\documentclass[
 aip,
article, 
reprint,
superscriptaddress,
nofootinbib,
nobibnotes,
 amsmath,amssymb,
floatfix,
]{revtex4-1}

\usepackage[version=3]{mhchem} 
\usepackage{multirow}
\usepackage{color}
\usepackage{graphicx}
\usepackage{dcolumn}
\usepackage{bm}
\usepackage{float} 
\usepackage{algorithm}
\usepackage[algo2e]{algorithm2e} 

\usepackage{cancel}
\usepackage{url}  
\usepackage{amssymb}
\usepackage{wasysym}
\usepackage{rotating}
\usepackage{mathrsfs}
\usepackage{subfigure}
\usepackage{psfrag}
\usepackage{verbatim}
\usepackage{amsfonts}
\usepackage{leftidx}
\usepackage{amsthm}
\usepackage{epstopdf}
\usepackage[version=3]{mhchem} 
\usepackage{array}
\usepackage{longtable}
\usepackage{booktabs}
\usepackage{siunitx}
\usepackage[export]{adjustbox}

\usepackage{mathptmx}

\usepackage{hyperref}
\hypersetup{colorlinks=true,citecolor=blue,urlcolor=blue}

\begin{document}

\preprint{AIP/123-QED}

\title{Peak thermoelectric power factor of holey silicon films}

\author{Jun Ma}
    \affiliation{Department of Mechanical Science and Engineering,  University of Illinois at Urbana-Champaign, Urbana, IL 61801, USA}
\author{Dhruv Gelda}
    \affiliation{Department of Mechanical Science and Engineering,  University of Illinois at Urbana-Champaign, Urbana, IL 61801, USA}
\author{Krishna V. Valavala}
    \affiliation{Department of Mechanical Science and Engineering,  University of Illinois at Urbana-Champaign, Urbana, IL 61801, USA}
\author{Sanjiv Sinha}
    \email{sanjiv@illinois.edu}
    \affiliation{Department of Mechanical Science and Engineering,  University of Illinois at Urbana-Champaign, Urbana, IL 61801, USA}
    \affiliation{Micro and Nanotechnology Laboratory, University of Illinois at Urbana-Champaign, Urbana, IL 61801, USA}

\date{\today}

\begin{abstract}

The thermoelectric properties of nanostructured silicon are not fully understood despite their initial promise. While the anomalously low thermal conductivity has attracted much work, the impact of nanostructuring on the power factor has mostly escaped attention. While initial reports did not find any significant changes to the power factor compared to the bulk, subsequent detailed measurements on {\it p}-type silicon nanowires showed a stark reduction in the Seebeck coefficient when compared to similarly doped bulk. The reduction is consistent with the disappearance of the phonon drag contribution, due to phonon boundary scattering. Here, we report measurements on a different nanostructure, holey silicon films, to test if similar loss of phonon drag can be  observed. By devising experiments where all properties are measured on the same sample, we show that though these films possess electrical conductivity  close to that in the bulk at comparable doping, they exhibit considerably smaller thermopower. The data are consistent with loss of phonon drag. At neck distances between 120 - 230 nm, the power factor at optimal doping is $\sim$ 50 percent that of the bulk. These insights are useful in the practical design of future thermoelectric devices based on nanostructured silicon.

\noindent \href{https://aip.scitation.org/doi/full/10.1063/5.0010254}{DOI:  https://doi.org/10.1063/5.0010254}
\end{abstract}

\maketitle

\section{Introduction}

In the last decade, silicon, one of the most abundant elements on earth, emerged as a promising candidate for thermoelectric applications in nanostructured forms, primarily due to experimental reports of high figures of merit~\cite{hochbaum2008enhanced,boukai2008silicon,tang2010holey,lim2015simultaneous,yu2010reduction,marconnet2012phonon,alaie2015thermal}. Since the enhancement was due to anomalously low thermal conductivity, subsequent research focused exclusively on understanding thermal transport phenomena in these structures. The other part of the figure of merit, the thermoelectric power factor $S^2\sigma$, where $S$ is the Seebeck coefficient and $\sigma$ is the electrical conductivity, has been relatively unexplored. Initial  work suggested that there is no change in power factor compared to the bulk, which at first glance appears consistent with the expectation that the nanostructures are large enough to not affect the density of electronic states. Apart from the fact that at the smallest features ($\sim$20 nm), structures are likely to suffer from surface charge depletion~\cite{bryant1984hydrogenic} and dopant segregation~\cite{fernandez2006surface}, bulk-like carrier transport is a reasonable expectation for high-quality fabrication.

Beyond carrier transport however, phonon transport is decidedly affected by nanostructuring. Therefore, a subtle question arises whether increased phonon boundary scattering that lowers thermal conductivity also affects the power factor through its relation to the Seebeck coefficient. The Seebeck effect in silicon arises not just due to the diffusion of charge under a temperature gradient but also due to drag on charge carriers from phonons that similarly diffuse under the temperature gradient. Herring~\cite{herring1954theory} pointed out the second contribution termed  phonon drag, which is known to be large in silicon at low temperatures. It is well known that phonon drag is reduced by boundary scattering of phonons, resulting in a scaling law proportional to sample size. Experimental verification of the law is often considered a vital proof of phonon drag. The contribution of phonon drag to the thermopower is difficult to pin down at room temperature through measurements or theory (see, for example, Ref.~[\onlinecite{sadhu2015quenched}]). However, extrapolating from low temperature behavior suggests that phonon drag should be strongly attenuated in nanostructures due to boundary scattering. The vital question is whether it is not already quenched in the bulk at room temperature at the high doping levels necessary in thermoelectric applications. 

In recent work~\cite{sadhu2015quenched}, we have shown that neither Umklapp scattering nor the presence of dopants quenches phonon drag completely in bulk silicon at room temperature. To the contrary, phonon drag contributes as much as a third of the Seebeck coefficient at room temperature even in silicon doped to $\sim 10^{19}$cm$^{-3}$. We deduced this by comparing the Seebeck coefficient of bulk silicon to that of similarly doped, rough surface nanowires. Analyzing the data for nanowires, we found a dramatic reduction in the Seebeck coefficient and attributed this to loss of phonon drag. These measurements call into question previous conclusions that the thermopower of silicon would be unaffected by nanostructuring. They further contradict decades-old measurements on bulk silicon~\cite{weber1991transport} that reached the conclusion that phonon drag is absent in heavily doped silicon. Subsequent {\it ab initio} calculations of phonon drag reached similar conclusion as our experiments. If the hypothesis is correct, we expect to observe similar reduction in thermopower in other silicon nanostructures. Here, we report measurements on {\it n}-type holey silicon films to test whether this is indeed the case. Periodic holey silicon~\cite{tang2010holey,lim2015simultaneous,yu2010reduction} refers to single crystalline silicon with a 2-D lattice of holes. The pitch is usually a few hundred to tens of nanometers and the hole diameter a fraction of the pitch. The limiting dimension for transport is the closest distance between holes, sometimes referred to as the neck distance. Periodic holey films are arguably the most important silicon nanostructures for practical application due to relative ease of integration.

We report data for 200 nm thick films with a fixed pitch of 350 nm and neck distances varying between 120 nm to 230 nm. The corresponding range of porosities are from 12 to 38 percent respectively. Compared with thermal-only characterization of such structures, detailed thermoelectric characterization in conjunction with thermal measurements is surprisingly rare even a decade after the initial discovery. This partly stems from the requirement for more sophisticated device fabrication~\cite{lim2015simultaneous,shi2003measuring} to perform different property measurements on the same sample. Here, we use an integrated platform that simplifies contact preparation and greatly facilitates simultaneous electrical and thermal measurements. The platform uses sinusoidal self-heating in doped silicon in combination with well established frequency-domain techniques ~\cite{lu20013omega} to measure the Seebeck coefficient, $S$ and the thermal conductivity, $k$. 


Our samples show a moderate reduction of thermal conductivity below the Casimir limit~\cite{casimir1938note} combined with bulk-like electrical conductivity. However, we observe a substantial decrease in the Seebeck coefficient when similarly conductive bulk silicon is used as the reference. The reduction is independent of limiting dimensions of holey films. We discuss the origins of this reduction in terms of lost phonon drag, and discuss issues with previous reports that concluded that phonon drag is negligible in highly doped silicon. We analyze the implications of the resultant loss in power factor. Finally, we provide quantitative guidance on a best-case power factor and identify the optimal doping. The paper is organized as follows. Section~\ref{sec:msmt} describes the measurement platform and the fabrication of devices. Section~\ref{sec:measurement_principle} discusses the measurement principles and Section~\ref{sec:data} the data obtained from our measurements. We model and discuss the implications of the data in Section~\ref{sec:discussion}. This work provides insight into practically achievable performances for nanostructured silicon thermoelectrics. 

\section{Measurement Device} \label{sec:msmt}

\begin{figure} [h]
   \includegraphics[width=1.0\linewidth]{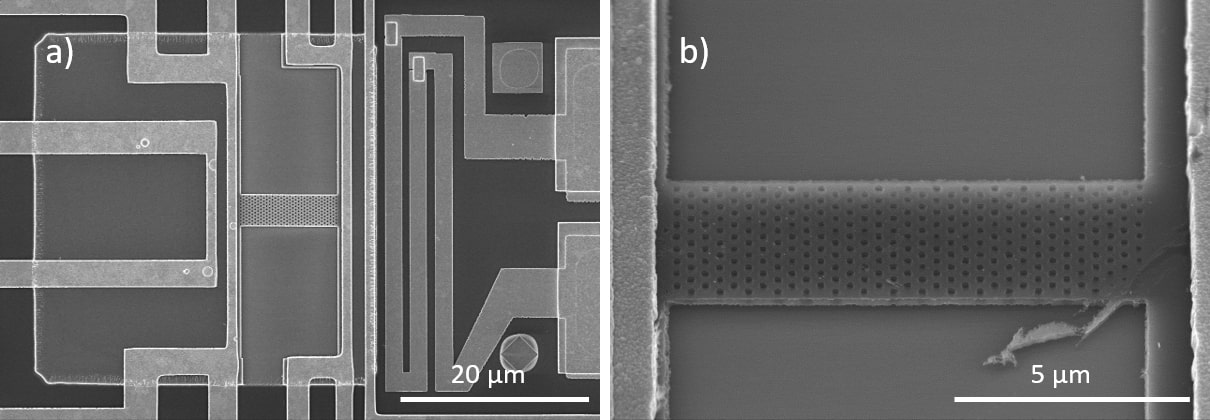}
   \caption 
   { \label{fig:SEM}(a) Top view SEM of a typical device. (b) Tilted view SEM showing suspended holey structures. 
 }
   \end{figure}

The measurement device follows previously published designs for in-plane measurements of the Seebeck coefficient~\cite{lee2009thermoelectric} but with  modifications to accommodate additional thermal characterization~\cite{lu20013omega}. A top-view SEM of a device is shown in Figure~\ref{fig:SEM}a. The holey silicon sample, doped in a phosphorous diffusion furnace and lithographically defined from an SOI device layer ($\sim$ 200 nm), is suspended in the middle and anchored by two pairs of metal lines. These metal lines consist of Cr/Au layers with thicknesses 5 nm and 300 nm respectively that are evaporated using an electron beam. They serve as electrodes in measuring the electrical conductivity and the Seebeck voltage. The inner pair, each of which is a resistor with four contact pads, also serve as resistance thermometers in Seebeck coefficient measurement. A serpentine metal line at one end of the sample is the heater that creates temperature gradients. Figure~\ref{fig:SEM}b shows a tilted view of the suspended holey silicon sample. 
 
In brief, microdevices were fabricated on SOI wafers with device layer $\sim$ 200 nm and buried oxide (BOX) $\sim$ 400 nm. The device layer, with varying barrier layer thickness (0-16 nm) was doped in a phosphorous diffusion furnace through pre-deposition at 950$^{\circ}$ C for 10 min followed by drive-in at 1100$^{\circ}$ C for 6 min. A 3$\%$ (diluted) HF removed any phosphosilicate glass and the barrier layer prior to the drive-in. These conditions in a bulk Si wafer without any barrier layer would result in a nearly constant doping concentration of $\sim$ 10$^{20}$ cm$^{-3}$ across the top 200 nm layer of silicon. Adding a barrier layer slows down the diffusion process and provides controlled doping depending on the thickness of the barrier layer. We can estimate the doping concentration profile, $C(x, t)$ during pre-deposition, where $x$ is the distance to oxide-silicon interface and $t$ is the pre-deposition time, by\cite{grove1967}

\begin{equation} 
\label{Eq:diffusion}
\frac{C(x, t)}{C_{O}}=\frac{2 m r}{m+r} \operatorname{erfc}\left[\frac{x_{O}}{2 \sqrt{D_{O} t}}+\frac{x}{2 \sqrt{D t}}\right],
\end{equation}

where $C_O$ is the saturation concentration (solubility) at the oxide surface, $m$ is the ratio of the concentrations on either side of the oxide-silicon interface (the segregation coefficient), $r=\sqrt{D_O/D}$, $D_O$ and $D$ are the phosphorus diffusivity in silicon oxide and silicon respectively, and $x_O$ is the thickness of the barrier layer. After removing any phosphosilicate glass and the barrier layer, the drive-in process re-distributes the dopant inside the device layer. Since the characteristic length $\sqrt{Dt}$ for drive-in process is comparable to layer thickness, the doping profile after drive-in is close to constant. Therefore, we can integrate the doping profile inside the device layer after pre-deposition to find the total dose and then assume a uniform value across the thickness to calculate the doping concentration after drive-in. We find the doping concentrations calculated in this manner for barrier layer thickness of 5nm, 10nm and 15nm to be 4.5$\times$ $10^{19}$ cm$^{-3}$, 2.3$\times$ $10^{19}$ cm$^{-3}$ and 1.0$\times$ $10^{19}$ cm$^{-3}$ respectively. 

After the doping process, a thin layer of SiO$_2$ ($\sim$ 75 nm) was deposited for insulation. Larger features such as contact pads and alignment markers were then defined by photolithography and smaller features such as electrodes, heaters and resistance thermometers were patterned using E-beam lithography. We used a 3$\%$ HF to remove the oxide between the two metallization steps so that the electrodes contacted the device layer. Next, the device went through a two-step etching process: a trimming step to avoid any cross-talk between different devices, and a patterning step to define the sample geometry and holey array. An STS ICP-RIE chamber was used for etching. Finally, a 3$\%$ HF undercut the BOX underneath the films that were then dried under supercritical CO$_2$. The schematic of the fabrication process is shown in \href{https://aip.scitation.org/doi/suppl/10.1063/5.0010254}{supplementary material} (Fig.~S1).

\section{Measurement Principle}\label{sec:measurement_principle}

We next describe the measurement principles in brief. For the property data reported here, the three property measurements were performed on the same sample unless a measurement could not be conducted, in which case the property is listed as NA.  The electrical resistance of each sample was first measured using 4-point probes at different temperatures to obtain the temperature coefficient of resistance, $dR/dT$. In the temperature range between 300K and 450K, we found $R$ to be linearly dependent on $T$, with $dR/dT$ a constant (see Fig.~S2 in \href{https://aip.scitation.org/doi/suppl/10.1063/5.0010254}{supplementary material}). 

The thermal conductivity was measured by heating the sample with a sinusoidal current $I$ (at 1$\omega$ frequency) applied across the outer electrodes using a lock-in amplifier\textquoteright s built-in sine output. A potentiometer with resistance much higher than the sample resistance was connected in series to mimic a current source. This current sets up a 2$\omega$ temperature oscillation due to Joule heating, which then creates a 3$\omega$ voltage $V_{3\omega}$ across the sample. The 3$\omega$ voltage is picked up by the lock-in amplifier with high dynamic reserve setting. The thermal conductivity $k$ can be determined by~\cite{lu20013omega}

\begin{equation} 
\label{Eq:thermal_conductivity}
k = \frac{I^3}{V_{3\omega}}\frac{4LR \frac{dR}{dT}}{\pi^4S\sqrt{1 + (2\omega\gamma)^2}},
\end{equation}

where $L$ is the length of the sample between inner electrode, $S$ is the cross sectional area, $\gamma=L^2/\pi^2\alpha$ is the thermal time constant of the sample, and $\alpha$ is the thermal diffusivity. Note that 2$\omega\gamma$ is equal to the ratio between the out-of-phase and in-phase signals of $V_{3\omega}$, i.e. $tan(\varphi)=2\omega\gamma$. In our measurement, $\gamma$ is on the order of $\mu$s and since the frequency used is $\sim$100 Hz, the out-of-phase signal is effectively zero. The linear relationship between 3$\omega$ voltage and 1$\omega$ current is shown in \href{https://aip.scitation.org/doi/suppl/10.1063/5.0010254}{supplementary material} (Fig.~S3). 

Finally, we measured the Seebeck coefficient using an approach similar to that described in Ref.~\onlinecite{lee2009thermoelectric} with the modification that we use frequency-domain sensing, as opposed to DC sensing, at both temperature sensors. A sinusoidal heating voltage $V_h$, supplied across the metallic heater, results a temperature field oscillating at frequency 2$\omega$. The amplitude of this oscillating temperature field at both ends of the sample are detected from the 3$\omega$ signals ~\cite{cahill1990thermal} across the two metallic RTD sensors. We measure the thermal transfer function by varying $V_h$ and extracting the proportionality constant $\alpha$ in $\Delta T = \alpha V_h^2$, where $\Delta T$ is the difference in amplitude of temperature oscillations between the two resistance sensors. We further measure the thermoelectric voltage using lock-ins at 2$\omega$ and similarly extract the proportionality constant $\beta$ in $V_{2\omega} = \beta V_h^2$. Note there is a 90$^{\circ}$ phase difference between the reference signal and V$_{2\omega}$. The Seebeck coefficient is $S = -\frac{\sqrt{2}\beta}{\alpha}$ where a factor of $\sqrt{2}$ appears since $V_{2\omega}$ is an RMS value while $\Delta T_{2\omega}$ is an amplitude. The measured $\alpha$ and $V_{2\omega}$ values are provided in the \href{https://aip.scitation.org/doi/suppl/10.1063/5.0010254}{supplementary material} (Figs.~S4-~S5).

\section{Thermoelectric Properties}\label{sec:data}

Table~\ref{tab:table4} summarizes data from our measurements. The sample label BL-x-S/H represents sample doping condition and structure characteristic (whether solid or holey). The number x is the thickness of barrier layer (SiO$_2$) during diffusion doping and S/H indicates solid or holey sample respectively. Due to fabrication yield issues, certain samples could not be fully measured. Their missing properties are marked `NA' in the table.

\begin{table*}[t]
\caption{\label{tab:table4}%
Summary of samples used in Seebeck and thermal measurement. The label read: BL for barrier layer, the number x for BL thickness in nm followed by S for solid or H for Holey. *Native oxide}
\begin{tabular}{c c c c c c}
\hline\hline
\textrm{Sample \#}&
\textrm{Est. doping}&
\textrm{Porosity}&
\textrm{Avg. D$_{neck}$}&
\textrm{S at 300 K}&
\textrm{k at 300 K}\\
\textrm{}&
\textrm{(cm$^{-3}$)}&
\textrm{(\%)}&
\textrm{(nm)}&
\textrm{($\mu$ V/K)}&
\textrm{(W/mK)}\\
\hline

BL-0*-S & $2\times 10^{20}$ & 0 & NA & NA & 37\\
BL-5-H & $8\times 10^{19}$ & 38 & 120 & -135 & NA\\
BL-10-H & $3.6\times 10^{19}$ & 38 & 120 & -186 & NA\\
BL-15-H1 & $1.3\times 10^{19}$ & 38 & 120 & -241 & 11\\
BL-15-H2 & $1.3\times 10^{19}$ & 12 & 230 & NA & 26\\
BL-16-S & $8\times 10^{18}$ & 0 & NA & -273 & NA\\

\hline
\end{tabular}
\end{table*}

\subsection{Electrical Conductivity}

\begin{figure}[h]
\includegraphics[width=1.0\linewidth, trim={1.4cm 0.6cm 2.7cm 1cm},clip]{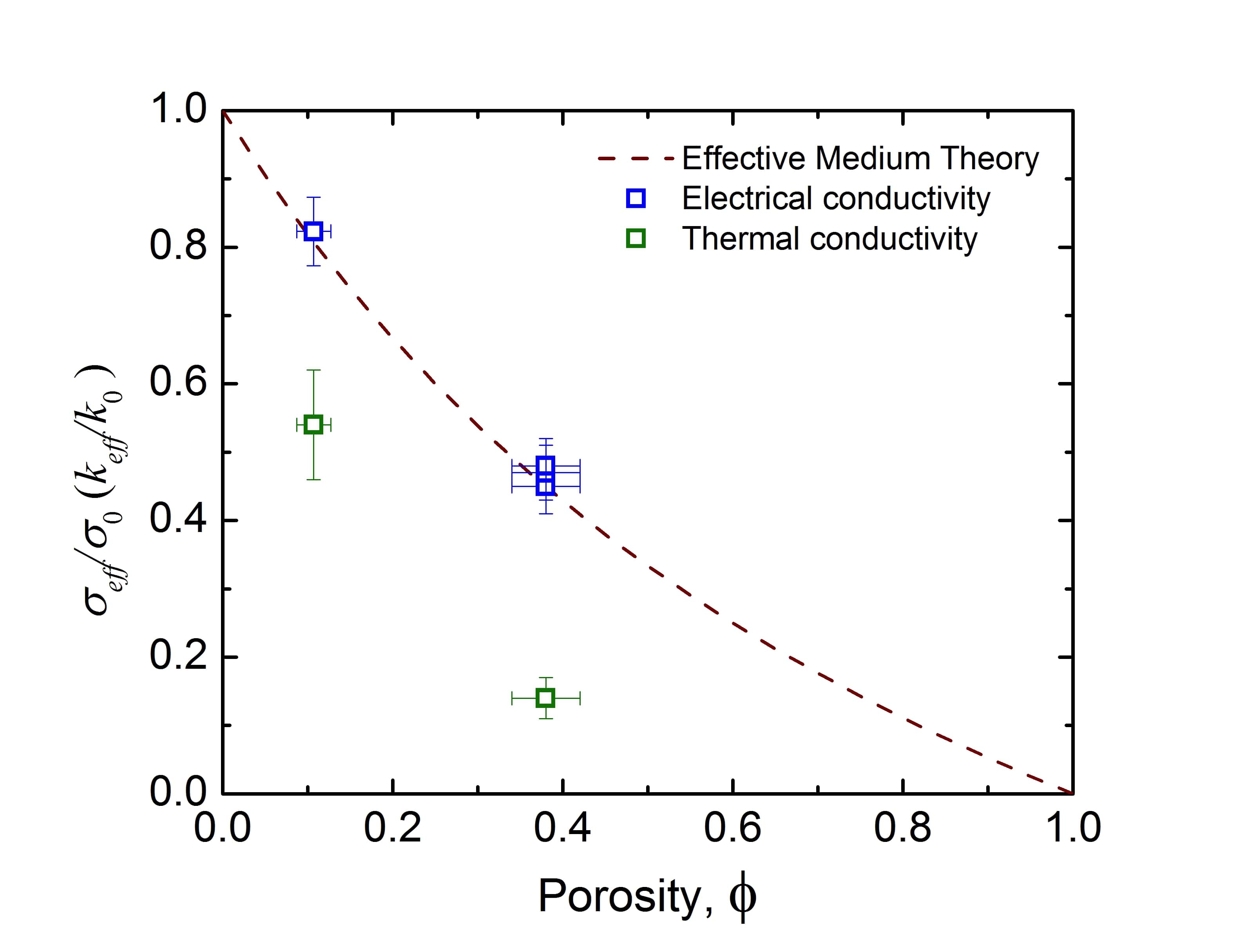}
\caption { \label{fig:effective} Ratio of the effective conductivity of the holey and solid thin films on the same chip plotted against porosity. The dashed curve represents the effective medium theory. Ratio of electrical conductivities follows the curve closely, while that of thermal conductivities falls significantly below the curve.}
\end{figure}

We now compare the electrical conductivity of holey silicon with that of solid thin films on the same chip. Both films experience identical doping conditions. Figure~\ref{fig:effective} plots the ratio of electrical conductivities as a function of porosity. The dashed curve represents the effective medium theory (EMT)~\cite{maxwell1881treatise,ma2014thermal} given by $\frac{\sigma_{eff}}{\sigma_{0}} = \frac{1-\phi}{1+\phi}$, where $\phi$ is the porosity. The excellent agreement between the data and the EMT curve suggests that carrier concentrations are indeed independent of the feature size of the holey films. The features range between $\sim$200 nm down to $\sim$120 nm in our samples. The plot also includes the ratio of thermal conductivities that is discussed next. At feature size $\sim$ 200 nm, temperatures above 300K and doping concentration $>$ 1 $\times$ $10^{18}$ cm$^{-3}$, we do not expect any significant surface charge depletion or dopant segregation. 

Based on prior work~\cite{bjork2009donor,kuo2013cmos}, we do not expect any effect on mobility at these film dimensions. Estimating the electron mobility from the measured electric conductivity and the theoretically estimated doping concentrations in Section II leads to mobility in excess of the value in the bulk, across all samples. This is expected given the limited accuracy of the doping concentrations from process modeling. Assuming instead bulk values of mobility and using the measured electrical conductivity of non-porous silicon thin films to extract carrier concentrations leads to the values listed in Table~\ref{tab:table4}. The carrier concentrations are in every case {\em greater} than those obtained theoretically from process modeling, within a factor of 2. This fact is important to the discussion on the Seebeck coefficient below. 

\subsection{Thermal Conductivity}

\begin{figure} [h]
\includegraphics[width=1.0\linewidth, trim={0.6cm 0.3cm 1cm 0.6cm},clip]{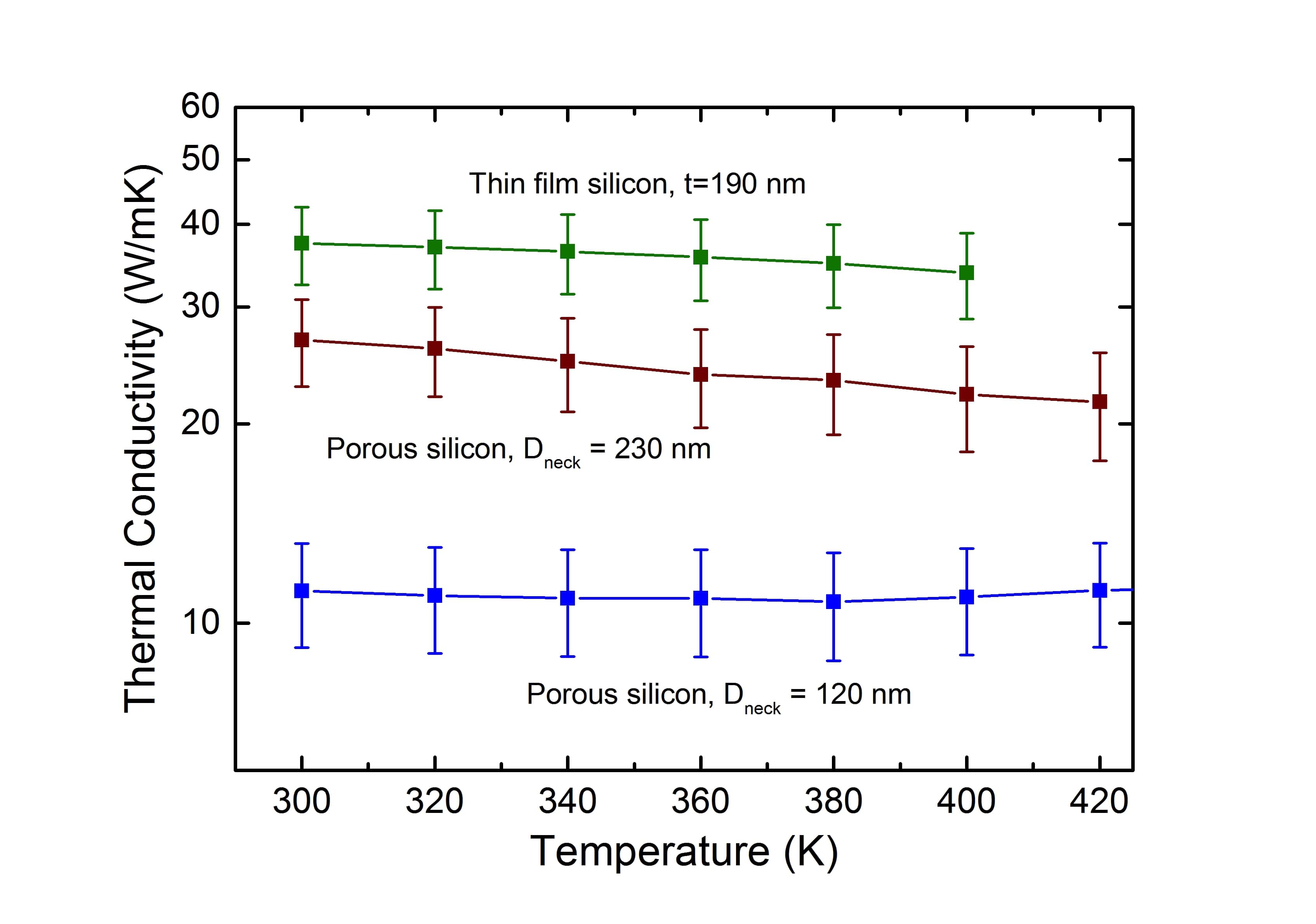}
\caption {\label{fig:thermal_conductivity}Measured thermal conductivity of one solid thin film and two holey silicon samples. The two films with larger limiting dimensions show a decreasing trend with temperature, that is less pronounced than bulk silicon. The thermal conductivity of the sample with smaller limiting dimension remains relatively flat with temperature.}
\end{figure}

Figure~\ref{fig:thermal_conductivity} plots the thermal conductivity of several films in the temperature range 300-420 K. The value of $k$ in the 190 nm thick solid film (reference) is 37 W/mK at 300 K. This is about 30-40 \% below the corresponding Casimir limit and is consistent with significant phonon electron scattering at the degenerate doping of 2$\times$10$^{20}$ cm$^{-3}$. Similar decrease in thermal conductivity has been observed~\cite{lim2015simultaneous} in Boron-doped silicon thin films at doping$\sim$4$\times$10$^{20}$ cm$^{-3}$. The addition of dopants is known to introduce phonon scattering due to mass differences, as well as due to free electrons. The total phonon relaxation time can be expressed using Matthiessen’s rule: $\tau^{-1} = \tau^{-1}_{U} + \tau^{-1}_{imp} + \tau^{-1}_{e} + \tau^{-1}_{b}$, where terms on the right hand side represent Umklapp, impurity, electron and boundary scattering respectively. Using expressions from the literature ~\cite{ma2012thermoelectric} for these scattering rates, the calculated values of $k$ are in agreement with the thermal conductivity of the solid film.

Figure~\ref{fig:thermal_conductivity} also shows the thermal conductivity of holey films with porosity 12 and 38\% respectively. In contrast to electrical conductivity, the ratio of the thermal conductivities of solid and porous films falls significantly below the EMT curve as shown in Figure~\ref{fig:effective}. Fitting the temperature dependent thermal conductivity using kinetic theory yields the mean free path to be 120 nm and 40 nm respectively. These are shorter than the neck distance in each case. Details of the fit are provided in the \href{https://aip.scitation.org/doi/suppl/10.1063/5.0010254}{supplementary material}.

\subsection{Seebeck Coefficient}

\begin{figure} [h]
\includegraphics[width=1.0\linewidth]{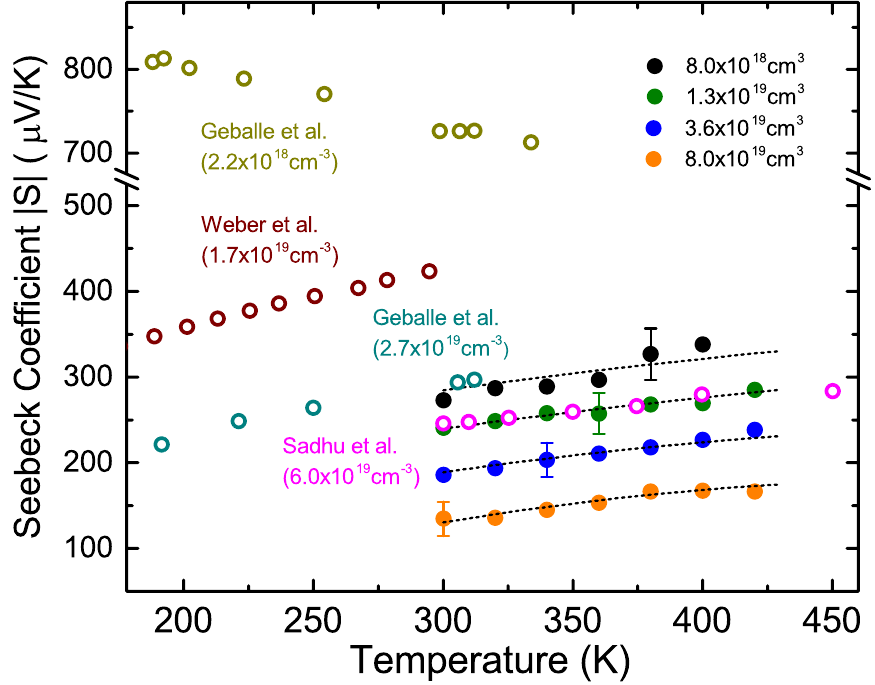}
\caption 
{ \label{fig:seebeck} Measured Seebeck coefficients at different doping concentrations as a function of temperature. Filled circles are measurements reported in this work. Black filled circles represent data for a solid thin film whereas the rest are for holey films. Literature data (Geballe et al.~\cite{geballe1955seebeck}, Weber et al.~\cite{weber1991transport}, and Sadhu et al.~\cite{sadhu2015quenched}) for bulk silicon is shown for comparison. The dashed curves are theoretical calculations of the diffusion part of $S$ assuming carrier concentrations listed in Table~\ref{tab:table4}.
}
\end{figure}

Figure~\ref{fig:seebeck} shows the measured $S$ for four different films in the temperature range 300-420 K. The reference sample is a solid thin film (filled black circles) doped to $8 \times 10^{18}$ cm$^{-3}$ whereas the rest are holey films with porosity 38\%. For comparison, we plot data for bulk silicon from the literature~\cite{geballe1955seebeck, weber1991transport} at similar doping . We also plot data from our previous measurements~\cite{sadhu2015quenched} on {\it n}-type bulk silicon doped to $6\times 10^{19}$ cm$^{-3}$ that were measured using the same frequency-domain technique as in this work. We have shown in prior work that measurements using the frequency-domain technique yield values for $S$ in bulk silicon in good agreement with classical work such as Ref.~[\onlinecite{geballe1955seebeck}], but with reduced uncertainties. The magnitude of $S$ decreases with increasing doping concentration, as expected from the increase in the Fermi level. All four samples exhibit an increasing trend with temperature. However, the measured $S$ is substantially smaller than that in the bulk at similar concentrations. We discuss the cause for this difference later in terms of a loss in phonon drag.

For the holey film labeled as BL-10-H in Table~\ref{tab:table4} with doping 3.6 $\times$ 10$^{19}$ cm$^{-3}$, we also measured a corresponding solid thin film sample with the same doping. Since we could obtain the measurement at 300 K before the failure of the solid film device, we do not include it in Figure~\ref{fig:seebeck}. The solid film had $S=-178 \mu V/K$ whereas the corresponding holey sample had $S=-186 \mu V/K$. The difference is within the measurement uncertainty. The absence of any difference in $S$ between solid and holey films is consistent with prior measurements~\cite{lim2015simultaneous}.  

In Fig.~\ref{fig:seebeck}, the dashed lines represent a theoretical fit of the diffusion component of the Seebeck coefficient ($S_d$) to the data. These calculations use the carrier concentrations listed in Table~\ref{tab:table4}. As mentioned previously, $S$ consists of two contributions, that from the diffusion of charge carriers, $S_d$ and that from the drag of phonons on charge carriers, $S_{ph}$. $S_d$ is proportional to the average energy transported by carriers with respect to the Fermi level, $E_F$ and can be written as 

\begin{equation} 
\label{Eq:effective1}
S_d = \frac{1}{eT}\Bigg(  \frac{\int{f'_0D(E)E(E-E_F)\tau dE}}{\int{f'_0D(E)E\tau dE}} \Bigg),
\end{equation}

where $f'_0$ is the differential Fermi distribution, $D(E)$ is electron density of states and $\tau$ is the electron relaxation time. For transport at room and higher temperatures, it is useful to introduce a scattering exponent, $r$ for the electron energy in the electron scattering time ($\sim$ $E^r$), such that Eq.~\ref{Eq:effective1} reduces to

\begin{equation} 
\label{Eq:effective2}
S_d = \frac{k_B}{e}\Bigg[ \frac{(r+2)F_{r+1}(\eta)}{(r+1)F_r(\eta)} - \eta\Bigg],
\end{equation}

where $\eta = E_F/k_BT$ is reduced Fermi energy and $F_j(\eta)$ is the $j^{th}$ order Fermi integral. The scattering exponent $r$ takes value 0 for acoustic scattering or boundary scattering and 2 for ionic impurity scattering. The dashed curves in Fig.~\ref{fig:seebeck} represent the diffusion component of each doping concentration calculated using Eq.~\ref{Eq:effective2}. We used $r$ as a fitting parameter to obtain best fit values of $r$ between 0.3-0.7. These values are well within the expected physical range of 0-2, suggesting that the measured $S$ can be explained entirely through $S_d$ and that $S_{ph}$ is negligible in our samples. 


\section{Discussion} \label{sec:discussion}

The reduction in $S$ of {\it n-}type holey silicon when compared to the bulk, are consistent with our observations in {\it p-}type silicon nanowires ~\cite{sadhu2015quenched}. The excellent fit of the theoretical $S_d$ to the data suggest that the measured $S$ is entirely due to the diffusion component. We have (limited) evidence that the presence of holes does not appear to affect $S$; rather $S$ is already reduced in solid thin films of similar thickness and doping. The reduction in $S$ points to a missing phonon drag component, $S_{ph}$ which must be far from negligible in the bulk even at high doping. This runs contrary to the decades-old notion that $S_{ph}$ is negligible at high doping. Recent {\t ab initio} calculations of phonon drag~\cite{zhou2015pnas} also found phonon drag to be substantial in bulk silicon at high doping. We next discuss possible explanations for this discrepancy between the older and more recent work.

The notion that phonon drag is suppressed in highly doped bulk samples stems partly from a conclusion reached by Weber and Gmelin~\cite{weber1991transport}, who revisited the classic measurements of Geballe and Hull~\cite{geballe1955seebeck} with modifications of their own. Weber and Gmelin considered the low temperature ($<$ 40 K) part of their data for degenerately doped Si and claimed $S\sim T$ in that range. Since Mott's formula for $S_d$ yields $S\sim T$, the authors concluded that only  was $S_d$ contributing to $S$. There are two subtle issues with this conclusion. First, drag may indeed vanish in their sample at T$<$40 K but the cause for the loss cannot be unambiguously determined to be from doping alone. Indeed, their 0.85x0.85 mm$^2$ sample was small enough for boundary scattering to dominate at low temperatures. Measurements on similarly sized samples by Geballe and Hull show clear evidence of boundary scattering and loss of phonon drag. Overlooking this, Weber and Gmelin further used the insight from Herring’s formula~\cite{herring1954theory} to conclude that drag was absent due to heavy doping. Herring’s formula under certain assumptions leads to the interpretation that $S_{ph} \sim 0$ in degenerate semiconductors. In the derivation provided by Herring, $S_{ph}=\beta u_0 \Lambda/{\mu_e T}$ where $\Lambda$ is the mean free path of phonons participating in drag, $\mu_e$ is the electron mobility, $u_0$ is the speed of sound and $\beta$ is the relative strength of electron-phonon scattering with respect to the ionic impurity scattering of electrons. Herring’s formula leads to $S_{ph}\sim0$ in degenerate semiconductors when (i) assuming $\beta\ll 1$, the so-called the Conwell-Weisskopf limit of ionic impurity scattering, (ii) underestimating the $\Lambda$ of long wavelength phonons and, (iii) not accounting for a phonon MFP spectrum $\Lambda(\omega)$ of drag phonons. We have discussed these issues in detail in Ref.~\onlinecite{sadhu2015quenched} where we present a revised formula in light of new insights such as the fact that phonon-electron rather than phonon-impurity scattering dominates at high doping. A more obvious second issue is that the $S\sim T$ fit is quite rudimentary; the fit (inset of Fig. 4, Ref.~[\onlinecite{weber1991transport}]) only applies over a narrow 5-15 K range. 

We now discuss the physics behind the loss of phonon drag in silicon nanostructures. At room temperature, phonons contributing to $S_{ph}$ are lower in frequency than those contributing to thermal transport. However, the spectrum of phonons contributing to drag is not confined stricty to the long wavelength limit. Herring's theory when used with detailed and empirically validated scattering rates for electrons and phonons taken from the modern semiconductor literature, provides a reasonable estimation of the rate of momentum exchange between electrons and phonons. Such calculations show that $\sim90\%$ of crystal momentum and thus phonon drag is contributed by phonon frequencies $\lesssim 1.2$ THz in non-degenerate silicon. In degenerate silicon, the magnitude of crystal momentum change is comparatively lower but the distribution is broadened with frequencies $\lesssim 2.3$ THz contributing to drag. At these frequencies, phonon specularity at a typical cleanroom fabricated surface is expected to be $<1$ (See Ref.~[\onlinecite{gelda2018specularity}], for example). Our calculations show~\cite{sadhu2015quenched} that at room temperature, boundary scattering affects $S_{ph}$ in silicon at limiting dimensions $<$ 10 $\mu$m, and that $S_{ph}$ is likely completely quenched when the limiting dimension $\sim$100 nm.


  
\begin{figure} [h]
\includegraphics[width=1.0\linewidth]{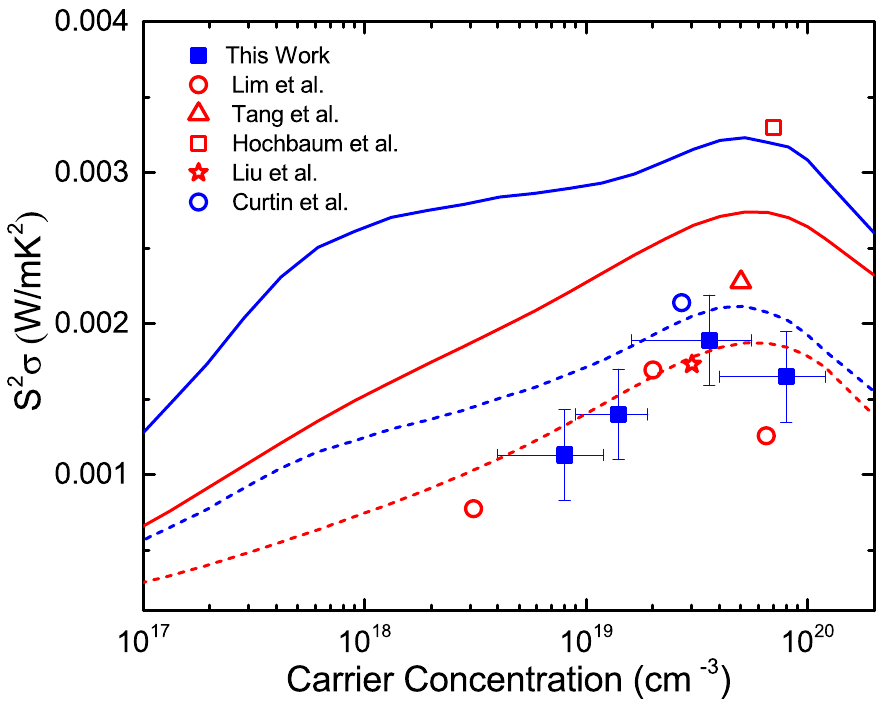}
\caption 
{ \label{fig:pf} Power factor comparison of {\it n}-type (blue) and {\it p}-type (red) silicon at 300K. The solid symbols are measurements from this work and the open symbols correspond to data from the literature (Lim et al.~\cite{lim2015simultaneous}, Tang et al.~\cite{tang2010holey}, Hochbaum et al.~\cite{hochbaum2008enhanced}, Liu et al.~\cite{liu2020thermoelectric}, and Curtin et al.~\cite{curtin2013field}). Solid and dashed curves represent calculations for bulk and nanostructured silicon respectively.
}
\end{figure}

The loss of $S_{ph}$ is very consequential to the overall performance of the material. At 300K, the highest $ZT$ in our measurements is 0.036, about a 4-fold increase from bulk silicon. The enhancement from the thermal conductivity reduction alone is about ten times, but more than half of this is compensated by the inferior power factor. Figure~\ref{fig:pf} further plots the power factor, $S^2\sigma$ against carrier concentration for {\it n}- (blue) and {\it p}-type (red) silicon at 300 K. The solid and dash curves represent calculations for the bulk and nanostructures respectively. The dashed curves only use S$_d$ from Eq.~\ref{Eq:effective2} for the Seebeck coefficient. We use results from our previous work on scattering rates for electrons and holes~\cite{ma2014thermal,seong2012modeling} to calculate the electrical conductivity. From our calculations, the optimal doping is in the range 4-6 $\times$ 10$^{19}$ cm$^{-3}$ across doping polarities. The difference between the solid and dashed curves at optimal doping represents $\sim$40\% drop in power factor in going from bulk to nanostructures, and is approximately the same for both {\it p}-type and {\it n}-type silicon. 

Figure~\ref{fig:pf} also plots room temperature data for comparison, including measurements reported in this paper (filled blue squares) as well as previously published work. The highest power factor across our measurements is 1.89 $\times$ 10$^{-3}$ W/mK$^2$ at 3.6 $\times$ 10$^{19}$ cm$^{-3}$, and is similar to that in {\it n}-type SOI~\cite{curtin2013field} doped at 1.7 $\times$ 10$^{19}$ cm$^{-3}$ (2.14 $\times$ 10$^{-3}$ W/mK$^2$, open blue circles) as well as that in gated silicon nanowires (1.86-2.28 $\times$ 10$^{-3}$ W/mK$^2$)~\cite{curtin2013field}. Data shown for {\it p}-type nanostructured silicon include nanowires~\cite{hochbaum2008enhanced}, holey silicon~\cite{tang2010holey} and ultrathin solid films~\cite{lim2015simultaneous}. As noted previously~\cite{sadhu2015quenched}, the power factor of nanowires in Ref.~[\onlinecite{hochbaum2008enhanced}] (open red square) is significantly higher than in other reported nanostructures. 

\begin{figure} [h]
\includegraphics[width=1.0\linewidth, trim={0cm 0 1.3cm 1cm},clip]{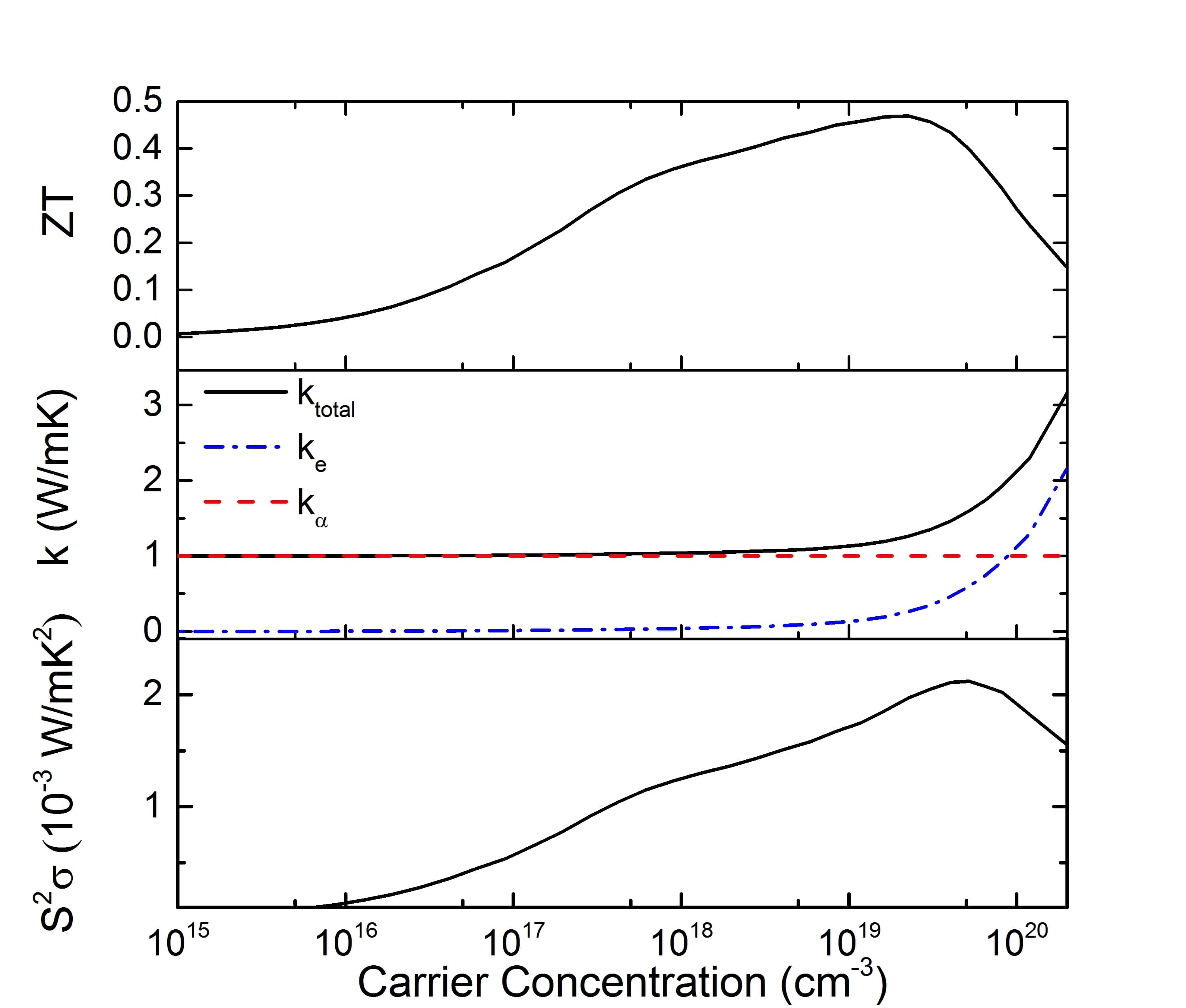}
\caption 
{ \label{fig:theoretical} Theoretical thermoelectric figure of merit $ZT$, minimum thermal conductivity $k$ and power factor as a function of carrier concentration in silicon nanostructure at 300K. The thermal conductivity includes electronic and lattice contribution with the latter taken at the amorphous limit.
}
\end{figure}

Figure~\ref{fig:theoretical} combines calculations for the power factor with best-case (lowest) estimates of thermal conductivity to plot a theoretically maximum  $ZT$ at 300 K. We assume the amorphous limit~\cite{cahill1992lower} as the minimum for the lattice contribution to thermal conductivity and the Wiedermann-Franz law to estimate the electronic contribution to the same. The two lower panels in Fig.~\ref{fig:theoretical} show the power factor and thermal conductivity. The electronic contribution to thermal conductivity overtakes the lattice at around 9 $\times$ 10$^{19}$ cm$^{-3}$ doping concentration. The upper panel in Fig.~\ref{fig:theoretical} shows the resulting $ZT$. Due to the addition of $k_e$, the optimal doping shifts slightly to the left to 2 $\times$ 10$^{19}$ cm$^{-3}$, where the corresponding $ZT$ peaks to $\sim$ 0.47. This represents a best-case $ZT$ that is experimentally challenging to reproduce given the contradicting requirements of small limiting dimensions and bulk-like electrical conductivity. 

\section{Conclusion}

In summary, we report thermoelectric measurements on holey silicon films using an integrated microdevice that combines frequency-domain measurements of thermal conductivity and the in-plane Seebeck coefficient. The complexity of fabrication is greatly reduced when compared to prior work utilizing measurement platforms specialized for one-dimensional nanostructures. At limiting dimension above 100 nm, we obtained bulk-like electronic conductivity in holey silicon, but find a degradation in $S$ compared to the bulk. The data fits the diffusion component of $S$ in Si, suggesting the loss of the phonon drag contribution. The observation is consistent with the conclusion from our previous report~\cite{sadhu2015quenched} on silicon nanowires. Accounting for the reduction in power factor, we identified the optimal doping for maximum performance to be 2 $\times$ 10$^{19}$ cm$^{-3}$ at which $ZT$ can be as high as $\sim$0.5. However, this maximum value requires limiting dimensions of $\sim$20 nm, where achieving bulk like electrical conductivity is challenging. The data and modeling presented here are useful in the design of future thermoelectric devices such as Peltier coolers based on nanostructured silicon~\cite{ren2017thermal, ren2018tsv}. Finally, we note that the loss of phonon drag is inconsequential at high temperatures ($>$ 600 K) where drag is absent even in the bulk due to Umklapp scattering. Thus, applications at elevated temperatures is unaffected by the results presented here and possibly constitute the best case for nanostructured silicon thermoelectrics.

\setcounter{secnumdepth}{0} 
\section{Supplementary Material}
See \href{https://aip.scitation.org/doi/suppl/10.1063/5.0010254}{supplementary material} for further details regarding the fabrication process and the measurement technique. 

\begin{acknowledgements}
The authors acknowledge support from the National Science Foundation through Grant No. NSF-CBET-17-06854 and the Air Force Office of Scientific Research through Grant No. AF FA9550-12-1-0073.
\end{acknowledgements}


\bibliography{paper_trim}

\end{document}